\documentclass[12pt]{article}

\def\appendix#1{
  \addtocounter{section}{1}
  \setcounter{equation}{0}
  \renewcommand{\thesection}{\Alph{section}}
  \section*{Appendices }
  \addcontentsline{toc}{section}{Appendix \thesection\ \ \ #1}
  }

\newcommand{\newsection}{    
\setcounter{equation}{0}
\section}

\def \ov {\over }
\def\bea{\begin{eqnarray}}
\def\eea{\end{eqnarray}}

\def\be{\begin{equation}}
\def\ee{\end{equation}}
\def\ba{\begin{eqnarray}}
\def\ea{\end{eqnarray}}

\def \bi{\bibitem}

\def\G{{\Gamma}}
\def\1{{{(1)}}}               \def\2{{{(2)}}}                      \def\3{{{(3)}}}

\font\mybb=msbm10 at 12pt

\def\bb#1{\hbox{\mybb#1}}

\def\Re {\bb{R}}

\def\Z {\bb{Z}}
\def\id{\protect{{1 \kern-.28em {\rm l}}}}

\def\nn{\nonumber}
\def \ov {\over }

\newcommand{\br}{\begin{array}}
\newcommand{\er}{\end{array}}
\newcommand{\beq}{\begin{equation}}
\newcommand{\eeq}{\end{equation}}

\newcommand{\sh}{\sinh}

\newcommand{\N}{{\cal N}}

\makeatletter
\@addtoreset{equation}{section}
\makeatother

\def\sH{{{\rm H}\!\!\!\!\hbox{\raise1pt\hbox{/}}\,}}
\def\sF{{{\rm F}\!\!\!\!\hskip.8pt\hbox{\raise1pt\hbox{/}}\,}}
\def\sh{{{\rm h}\!\!\!/\hskip.4pt }}
\def\sf{{{\rm f}\!\!\!\hskip.8pt\hbox{\raise1pt\hbox{/}}}}

\begin{document}

\begin{titlepage}

\begin{flushright}
hep-th/0206195
\end{flushright}
\bigskip \bigskip 
\begin{center}
{\LARGE \bf Supergravity pp-wave solutions with 28 and 24 supercharges}
\end{center}
\bigskip       
\centerline{{\large \bf Iosif Bena and Radu Roiban}}  
\medskip 
\centerline{ Department of Physics } 
\centerline{University of California} 
\centerline{Santa Barbara, CA  93106}
\medskip 
\centerline{{\rm iosif, radu@vulcan.physics.ucsb.edu} }

\vskip1.5truecm

\abstract{ We conduct an exhaustive search for solutions of IIA and 
IIB supergravity with augmented supersymmetry. We find a two-parameter 
family of IIB solutions preserving 28 supercharges, 
as well as several other IIA and IIB families of solutions with 24 
supercharges. Given the simplicity of the pp-wave solution, the 
algorithm described here represents a systematic way of classifying all such
 solutions with augmented supersymmetry. By T-dualizing some of 
these solutions we obtain exact non-pp wave supergravity solutions 
(with 8 or 16 supercharges), which can be interpreted as perturbations of 
the AdS-CFT correspondence with irrelevant operators.  }

\end{titlepage}


\newsection{Introduction}

Plane waves are among the simplest solutions of the supergravity equations of 
motion. Due to the existence of a null Killing field, they are also 
solutions of string theory to all orders in the sigma model perturbation
theory \cite{HOST, AMKL}. 

Besides the three well-known supergravity solutions with 32 
supercharges ($AdS_{4,5,7}\times S^{7,5,4}$), it is possible to construct 
two more \cite{pp,f1,f2}. Even though originally
these solutions were constructed by solving the equations of motion, it 
later turned
out that they can be obtained as Penrose-G\"uven limits \cite{penrose}
of the former. One of these solutions \cite{pp,f1} is a pp-wave in 11 dimensional
supergravity, has a nonzero four-form field strength $F_4$, and 
is the Penrose-G\"uven limit of both $AdS_4\times S^7$ and  
$AdS_7\times S^4$. The other solution \cite{f2} is a pp-wave in 10 
dimensional type IIB supergravity, has a nonzero self-dual five-form 
field strength $F_5$, and is the Penrose-G\"uven limit of 
$AdS_5\times S^5$. These important observations provided the link between 
plane wave solutions of supergravity equations of motion and the 
AdS/CFT correspondence. Thus, string theory in the plane wave geometry is dual to 
a sector with large R-charge on the gauge theory side \cite{bmn}.

The ensuing burst of interest in plane wave geometries 
prompted the construction of solutions \cite{c1,c2,h1} generalizing
the original ones and preserving more supersymmetries than the standard
16 of any plane wave.

The plane wave geometry seems simple enough to attempt a classification
of these augmented supersymmetry solutions. In this paper we perform 
this analysis for type IIB and IIA supergravity with the surprising result 
that in the type IIB theory there exist solutions preserving 28 
supercharges. 
Our method is powerful enough to allow the classification 
of all solutions with 24 supercharges as well. We construct a fairly large 
number of them, both in type IIA and type IIB supergravity. Even though we do not 
prove here that our analysis exhausts all these solutions, we believe it is quite 
likely that it does. In the type IIA theory
we also find solutions preserving $(p,q),\,p\ne q$ supercharges. Four dimensional 
solutions with this property were also constructed in \cite{s}.

In the presence of general form fields, the dilatino variation is 
proportional to the contraction of these forms with the Dirac $\G$ 
matrices acting on the supersymmetry parameter $\epsilon$. 
Therefore, in order to obtain more preserved supersymmetries,
one needs the $\G$ matrices to combine into commuting projectors.
In order for this to happen one needs to turn on appropriate forms with 
appropriate coefficients. 

If the dilatino variation takes the form
\be
\delta \lambda = M (\G_0\G_-)(1+M_1) (1+M_2) \epsilon,
\label{dilatino}
\ee
where $M$ is a matrix, $M_1$ and $M_2$ are independent, commuting and 
unipotent ($M_i^2=1$) combinations of $\G$ matrices, 
each of the three projectors will annihilate half of the 
spinors it acts upon. Since we assumed them to be independent and commuting
they will annihilate {\em different} sets of spinors and thus the right 
hand side of (\ref{dilatino}) will vanish for $16+8+4=28$ spinors. If 
instead of three projectors we only have two, then only $16+8=24$ 
spinors give a zero dilatino variation.

Once we have these candidates for Killing spinors, the next step is
to test whether the gravitino supersymmetry variation vanishes. 
For plane wave would-be solutions this completely fixes the 
metric, as well as the dependence of the spinors on the coordinates. In 
some cases all the  24 or 28 spinors give a zero gravitino variation, so 
they are Killing spinors. In other cases, the number of Killing spinors 
is smaller.

In the next section of this paper we describe the pp-wave geometry and the 
form of the dilatino and gravitino supersymmetry variations. We then explore
the types and combinations of form fields that can be turned on in order 
for projectors to appear in these equations. Then, we use the dilatino 
variation to make a number of educated guesses for solutions with 
enhanced supersymmetry, both in type IIA and in type IIB supergravity.
In section 4 we test these ansatze against the gravitino variation, and find the 
full solutions.

We first describe two families of type IIB backgrounds with 
$28$ Killing spinors. These backgrounds have nonvanishing self-dual $F_5$ 
flux, as well as nonzero RR or NSNS three forms in particular combinations. 
The relative strength of the five form and RR or NSNS three form is a 
free parameter, so each of the two solutions is in fact a one parameter 
family. \footnote{Of course these two solutions are related by S-duality, and are just
the end points of an entire family of solutions generated by rotating $F_3$ and $H_3$ 
into each other via S-duality. This gives in the end a 2-parameter family of 28 supercharge solutions.} 

We then list the other type IIA and type IIB backgrounds with 
more than $16$ Killing spinors which we obtain by this procedure. 
We list solutions with 24 supercharges involving $F_3 +F_5$, $H_3+F_5$, 
$F_3+ H_3+ F_5$,  $F_4 + H_3$, $F_4 + F_2$, $H_3+ F_2$, as well as solutions preserving 
chiral supersymmetry.

Some of the solutions we analyze have some Killing spinors independent
of the coordinate along the direction of propagation of the wave. Thus,
it is possible to T-dualize along this direction and still have a solution 
preserving some supersymmetry. We find that the dual geometries can be 
interpreted as arising from smeared strings or D-branes deformed with 
transverse fluxes, and explain them in light of the AdS-CFT correspondence.
In the process we construct exact nonsingular flows from  brane near-horizon 
geometries in the IR to certain non-trivial geometries in the UV.  
The results are described in section 5. 

It is also interesting to ask what is the highest number of supersymmetries 
than can be preserved by a pp-wave background in type II theories. To 
obtain 32 supercharges one needs the dilatino variation to vanish, in order
to impose no constraints on the supersymmetry parameters. Thus the only 
form field we can have is the type IIB self-dual five-form. The maximally 
supersymmetric pp-wave background obtained in \cite{f2} is the only 
such solution.

A solution preserving  30 supercharges would have a dilatino variation  
containing a product of four independent projectors. As we will show in 
section \ref{general}, it is not possible to combine the
fields of IIA and IIB supergravity to form so many projectors. Thus, besides 
the maximally supersymmetric solution of type IIB supergravity, the 
solutions with 28 supercharges described here have the largest possible 
amount of supersymmetry one can obtain in a pp-wave background in 
10 dimensions\footnote{It would 
be interesting to see if the methods we use for finding pp-waves solutions 
with augmented supersymmetry (combining the forms to form projectors) 
can be used to find M-theory or lower dimensional supergravity solutions with 
28 supersymmetries.}.


\section{Supersymmetries and projectors\label{susyproj}}

${~~~\,}$ In this section we will describe in detail a general way of 
constructing wave solutions of the supergravity equations of motion 
with enhanced supersymmetry. 

As it is known, the metric and forms of a pp-wave are quite simple, yet 
nontrivial. We choose a metric of the form:
\be
ds^2=-2dx^+dx^- - A_{ab}(x^+)z^az^b(dx^+)^2+(dz^a)^2~~,
\label{4.1}
\ee
and the only nonzero component of the field strengths of the RR and NSNS 
fields is $F_{+i_1...i_p}(x^+)$. Because they only depend on $x^+$, the forms 
satisfy the equations of motion and Bianchi identities by construction.

Choosing ($\eta_{+-}=-1$), the vielbeine are: 
\bea
&&e^+=dx^+~~~~~~~~e^a=(d\phi,\,dx^i,\,dy^i)\equiv dz^a\nn\\
&&e^-=dx^-+{1\ov 2}A_{ab}(x^+)z^az^bdx^+,
\eea
and the spin connection (defined by $de^A+\omega^A{}_B\wedge e^B=0$) is:
\be
\omega^{-c}=A_{cb}(x^+)z^bdx^+.
\label{spincon}
\ee
The supercovariant derivative is therefore given by:
\be
\nabla_i=\partial_i~~~~~~~~\nabla_-=\partial_-~~~~~~~~
\nabla_+=\partial_++{1\ov 2}A_{ab}(x^+)z^b\Gamma_-\Gamma_a,
\ee
and the Ricci tensor is just
\be
R_{++}=A^a{}_a(x^+).
\ee
Thus, the only equation of motion our backgrounds have to satisfy is
\be
R_{++}=A^a{}_a(x^+)= {1\ov 2}
\sum_p{{1\over p!}F_{+i_1...i_p } F^{+i_1...i_p }},
\label{eom++}
\ee
where $ F_{+i_1...i_p } $ are the field strengths of the various RR and 
NSNS $p-$forms present and self-dual fields enter only once.

We will use the conventions of  \cite{GRAN} for the type II supersymmetry 
transformation rules. In these conventions we will work with two 
Dirac spinors (thus, all Dirac matrices will be 32-dimensional) obeying 
appropriate chirality conditions and forming a
2-dimensional representation of an auxiliary $SL(2,\Re)$. 
Defining $\sF{}_{(n)}={1\ov n!}\Gamma^{N_1\dots N_n}
F_{N_1\dots N_n}$, the supersymmetry transformations are:

\noindent $\bullet~{\rm type}~~IIA:$
\bea
\delta\lambda\!\!&=&\!\!{1\ov 2}\Gamma^M\partial_M\phi\epsilon
-{1\ov 4}\sH\sigma^3\epsilon
+{1\ov 2}e^\phi\left[5 F_{(0)}\sigma^1+3\sF{}_{(2)}(i\sigma^2)
+\sF{}'_{(4)}\,\sigma^1\right]\epsilon\label{IIAsusy}\\
\delta\Psi_M\!\!&=& \!\!\nabla_M
\epsilon-{1\ov 8}\Gamma^{NP}H_{MNP}\sigma^3\epsilon+
{1\ov 8}e^\phi
\left[F_{(0)}\Gamma_M\sigma^1+\sF{}_{(2)}\Gamma_M(i\sigma^2)
+\sF{}'_{(4)}\Gamma_M\,\sigma^1
\right]\epsilon\nn
\eea
\noindent $\bullet~{\rm type}~~IIB$
\bea
\delta\lambda\!\!&=&\!\!{1\ov 2}\Gamma^M\partial_M\phi\epsilon
-{1\ov 4}\sH\sigma^3\epsilon
-{1\ov 2}e^\phi\left[\sF{}_{(1)}(i\sigma^2)\epsilon+
{1\ov 2}\sF{}'_{(3)}\sigma^1\epsilon\right]  \label{IIBsusy}  \\
\delta\Psi_M\!\!&=& \!\!\nabla_M\epsilon
-{1\ov 8}\Gamma^{NP}H_{MNP}\sigma^3\epsilon
+{e^\phi\ov 8}
\left[\sF{}_{(1)}\Gamma_M(i\sigma^2)+\sF{}'_{(3)}
\Gamma_M\sigma^1
+{1\ov 2}\sF{}'_{(5)}\Gamma_M(i\sigma^2)\right]
\epsilon\nn
\eea
with the modified field strengths $F'$ given by:
\be
F'_{(3)}=F_{(3)}-C\,H_{(3)}~~~~~~
F'_{(5)}=F_{(5)}- H_{(3)}\wedge C_{(2)}~~~~~F'_{(4)}=F_{(4)}- H_{(3)}\wedge C_{(1)}
\ee

Preserved supersymmetry appears in the form of spinors that are
annihilated by a set of projectors when the above transformations are 
evaluated on solutions to the equations of motion. Thus, a classification of
all possible solutions preserving some supersymmetry becomes a three-step
process. The first step requires a classification of projectors that can be built 
out of supergravity fields in the dilatino transformation rule. The next step
requires checking whether these field configurations are compatible with 
the gravitino supersymmetry transformation (Killing spinor equation)
and the third step involves checking whether the equations of motion are 
satisfied.

The first step in the procedure outlined above can be performed 
in quite some generality. In the notation we are using here a generic 
projector looks like
\be
P={1\ov 2}(1+\Gamma\otimes\sigma)~~~~~~~~
\left(\Gamma\otimes\sigma\right)^2=1
\label{projector}
\ee
where $\Gamma$ is some combination of Dirac matrices and $\sigma$
is one of the $gl(2,\Re)$ generators. We will loosely refer to the $\sigma$-dependence
of various terms as their $gl(2)$ structure. Half of the eigenvalues of  such a projector 
vanish. Thus, one such projector will preserve one half of the 
supersymmetries. The only way to find more preserved supersymmetries is
to have the dilatino variation be proportional to a product of commuting 
projectors. This observation  allows us to find the maximum number 
of supersymmetries that can be preserved by a solution of the 
equations of motion which has a nontrivial supersymmetry 
transformation of the dilatino\footnote{For the maximally supersymmetric wave 
each term in the dilatino variation vanishes separately}.

Due to the fact that we are considering wave solutions, each term in the 
dilatino variation is proportional to the Dirac matrix pointing along the 
(null) direction of propagation of the wave (this direction will be 
denoted by $x^+$). This matrix is proportional to a projector 
(\ref{projector}) in which $\sigma=1$. It is easy to see that this 
projector commutes with any other projector that can be constructed 
from the remaining Dirac matrices appearing in the supersymmetry 
transformation rules. Thus a wave solution will always preserve sixteen 
supercharges.

In the next two sections we will follow the steps outlined above. We will 
begin by describing several field configurations that factorize the dilatino 
variation into projectors. These field configurations have the potential of 
producing wave solutions preserving $28$ supercharges. 
We will then proceed in section \ref{eom} to analyze the Killing spinor 
equation and the equations of motion.


\section{Potential solutions\label{secdilatino}}

${~~~\,}$ Given the simplicity of the wave metric and the fact that all field 
strengths carry one null index, it is easy to find field configurations such 
that the dilatino transformation is proportional to a product of commuting 
projectors. 

We will begin with the type IIB supergravity. It will be argued in section 
\ref{general} that the dilaton and the axion cannot have nontrivial values
if more that $16$ supercharges are to be preserved. Thus, we will look for 
field configurations involving only the 3-form field strengths.

We will first discuss potential solutions with either one of $H_{(3)}$ or 
$F'_{(3)}$ nonvanishing. It is very easy to see that, after factorizing the 
Dirac matrix pointing along the direction of propagation of the wave, 
both $H_{(3)}$ and  $F'_{(3)}$ will contribute two Dirac matrices that must 
be further combined in projectors. Since for the time being we are 
considering only one type of field, the $gl(2)$ component of the 
supersymmetry transformation rule will factorize. The only possibility is 
then to find projectors constructed out of four Dirac matrices.  It turns out 
to be possible to have
\be
\delta\lambda\sim \Gamma_-
\left(1-\beta\Gamma_{1234}\right)
\left(1-\gamma\Gamma_{1256}\right)\epsilon~~,~~~~~~~~
\beta^2=\gamma^2=1
\label{p1}
\ee
which vanishes for $28$ different spinors. The field configuration realizing 
this setup is the following:
\be
H_{+12}=\beta H_{+34}=\gamma H_{+56}=
\alpha \beta \gamma H_{+78}=f(x^+)~~~~~~~~
\Gamma_{-1}\epsilon=\alpha\epsilon~~~~~~~~\alpha^2=1~~.
\label{fc1}
\ee
Here $f(x^+)$ is for the time being an arbitrary function of $x^+$ while
$\G_{-1}$ is the 10-dimensional chirality operator. As stated 
in the beginning, we are free to replace $H$ with $F'$. This function will be 
fixed in the next section using the gravitino variation as well as the equations 
of motion.

To show that this field configuration indeed reproduces (\ref{p1}) we
need to make use of the fact that both supersymmetry parameters have the 
same chirality. After pulling out $\G_-\G_{12}$ as common factor 
the dilatino variation becomes
\bea
\delta\lambda&=&{f(x^+)\ov 4}
\Gamma_-\Gamma_{12}\otimes\sigma^l
\left[1-\beta\Gamma_{1234}-\gamma\Gamma_{1256}
-\alpha\beta\gamma\Gamma_{1278}\right]
\epsilon\nn\\
&=&{f(x^+)\ov 4}
\Gamma_-\Gamma_{12}\otimes\sigma^l
\left[1-\beta \Gamma_{1234}-\gamma\Gamma_{1256}
-\alpha\beta\gamma\Gamma_{3456}\Gamma_{-1}\right]\epsilon\nn\\
&=&{f(x^+)\ov 4}
\Gamma_-\Gamma_{12}\otimes\sigma^l
\left(1-\beta\Gamma_{1234}\right)
\left(1-\gamma\Gamma_{1256}\right)\epsilon
~~~~~~~~~~~~~~~~l=1,3
\eea
Here we used the definition of the chirality operator 
$\Gamma_{-1}=-{1\ov 2}[\Gamma_{+},\,\Gamma_-]\Gamma_{12345678}$ 
\footnote{$\Gamma_{\pm}={1\ov \sqrt{2}}(\Gamma_0\pm \Gamma_9)$; 
$\{\Gamma_+,\,\Gamma_-\}=-2$}
and the definition of $\alpha$ in equation (\ref{fc1}). The choices $l=1$ and 
$l=3$ correspond to having a nontrivial RR 3-form and NSNS 3-form, 
respectively. S-duality continuously interpolates between these two solutions.

Next we discuss a possible solution of type $IIB$ supergravity which 
preserves 28 supercharges, contains both $F_{(3)}$ and $H_{(3)}$ and
is not S-dual to the solutions considered above (neither $F_{(3)}$ 
nor $H_{(3)}$ can be S-dualized away). It is clear that, after pulling out 
a common factor, some terms
will be left with the identity operator as their $gl(2)$ component while others
will have $(i\sigma^2)$. Since $(i\sigma^2)$ can appear in a projector only 
tensored with two or three Dirac matrices, it is easy to see that a possible 
combination of projection operators is:
\be
\delta\lambda\sim \Gamma_-\left(1-\beta\Gamma_{14}(i\sigma^2)\right)
\left(1-\gamma\Gamma_{23}(i\sigma^2)\right)\epsilon~~~,~~~~~
\beta^2=\gamma^2=1~~.
\ee
The field configuration producing this dilatino variation is:
\be
\gamma H_{+13}=-\beta H_{+24}=F'_{+12}=\beta\gamma F'_{+34}=f(x^+)~~.
\label{fc2}
\ee
As before, $f(x^+)$ is for the time being arbitrary and will be determined by 
the gravitino variation and equations of motion. One can in principle 
construct these field strengths from several different potentials. However, 
we choose the gauge in which the potentials do not carry the $+$ index.
The reason for this gauge choice is to make sure that the modified 5-form 
field strength remains trivial. As promised, the dilatino variation is:
\bea
\delta\lambda&=&-{f(x^+)\ov 4}\Gamma_-
\left[\left(\gamma\Gamma_{13}-\beta\Gamma_{24}\right)\sigma^3+
\left(\Gamma_{12}+\beta\gamma\Gamma_{34}\right)\sigma^1
\right]\epsilon\nn\\
&=&-{f(x^+)\ov 4}\Gamma_-\Gamma_{12}\sigma^1
\left(1-\beta\Gamma_{14}(i\sigma^2)\right)
\left(1-\gamma\Gamma_{23}(i\sigma^2)\right)\epsilon~~.
\eea
We will show in section 6 that, up to a relabeling of coordinates, 
the field configurations described above are the only ones that lead to a product 
of three projectors (two if one ignores $\Gamma_-$) in the dilatino 
supersymmetry transformation rule.


We now turn to possible solutions of type IIA supergravity. As in 
type IIB supergravity, any wave solution preserving more that $16$ 
supercharges has a trivial dilaton.
Even with this simplification, the situation is substantially more complicated
than in type IIB theory since there are three  different types of fields 
contributing to the dilatino transformation rule. Deferring the detailed 
analysis to section \ref{general}, we present here several examples. 

The only possible (up to relabeling and reshuffling 
of terms) projector that could preserve 28 supercharges is:
\be
\delta\lambda\sim \Gamma_-\left(1+\beta\Gamma_{148}(i\sigma^2)\right)
\left(1+\gamma\Gamma_{245}(i\sigma^2)\right) ~~,
\label{IIA28-1}
\ee
and the field 
configuration generating it is: 
\be
{1\ov 2}H_{+12}={1\ov 2}\beta\gamma H_{+58}=-\beta F_{+248}=
\gamma F_{+145}=f(x^+)~~.
\label{IIA28-field}
\ee
As before $f(x^+)$ is an arbitrary function to be determined by the Killing spinor 
equation and equations of motion. The dilatino variation generated by this 
field configuration is indeed proportional to (\ref{IIA28-1})
\be
\delta\lambda=\Gamma_-\Gamma_{12}\otimes\sigma^3
\left(1+\beta\Gamma_{148}\otimes(i\sigma^2)\right)
\left(1+\gamma\Gamma_{245}\otimes(i\sigma^2)\right)
\ee
It is possible to add a further projector to the product above. However, 
this requires use of $\Gamma_{-1}$ and thus it enhances supersymmetry 
only in the right-handed sector while breaking it in the left-handed sector.

Finding solutions preserving 24 supercharges is also easy in this approach.
The projector:
\be
\Gamma_-\left(1+\beta \Gamma_{1}\otimes\sigma^1\right)
\ee
can appear in a solution with nonzero $H_{(3)}$ and $F_{(2)}$:
\be
{3}F_{+1}=-{1\ov 2}H_{+12}=f(x^+), 
\label{f2h3}
\ee
This gives the dilatino variation:
\be
\delta\lambda={f(x^+)\ov 2}\Gamma_-\left(\Gamma_{12}\sigma^3+
\Gamma_1(i\sigma^2)\right)
=-{f(x^+)\ov 2}\Gamma_1(i\sigma^2)\Gamma_-
\left(1-\Gamma_{2}\sigma^1\right)~~,
\label{dilf2h3}
\ee
which contains the projector promised above.

Another example of potential solutions preserving 24 supercharges is built 
on the projector:
\be
\Gamma_-\left(1+\Gamma_{1234}\sigma^3\right)~~.
\ee
The field configuration that can generate this projector contains $F_{(4)}$ and $F_{(2)}$:
\be
3F_{+1}=F_{+234}=f(x^+)~~.
\label{f2f4}
\ee 
This leads to the dilatino variation
\be
\delta\lambda={f(x^+)\ov 2}\Gamma_-\left(\Gamma_1(i\sigma^2)
+\Gamma_{234}\sigma^1\right)=-{f(x^+)\ov 2}\Gamma_1(i\sigma^2)
\Gamma_-\left(1
+\Gamma_{1234}\sigma^3\right)~~.
\label{dilf2f4}
\ee

To summarize, we have described how the study of the dilatino variation can yield 
field configurations that have the potential of 
preserving large amounts of supersymmetry. The final word in this matter belongs 
however to the Killing spinor equation and the supergravity 
equation of motion. We proceed with their analysis, thus completing 
the second and third steps of the program 
outlined in section \ref{susyproj}.


\section{Gravitino variation and equations of motion.\label{eom}}

\subsection{Generalities}

The strategy for solving the Killing spinor equations in plane wave backgrounds 
was discussed in some detail in \cite{f1,c1}. Here we will go beyond their analysis and cast these 
equations in a form suitable for the setup discussed in the previous sections. 

The generic structure of the gravitino transformation is
\be
\delta\Psi_M=\nabla_M\epsilon+\Omega_M(x^+)\epsilon
\ee
where $\Omega_M(x^+)$ is the torsion part of the spin connection
and represents the contribution of the various form fields. If all RR fields 
vanish then $\Omega_M$ is just the standard torsion induced by the
NSNS 3-form field strength. It is not hard to see from the
gravitino variations (\ref{IIAsusy}) and (\ref{IIBsusy}) that
$\Omega_i(x^+)$ is proportional to $\Gamma_-$. Therefore 
\be
\Gamma_-\Omega_i(x^+)=\Omega_i(x^+)\Gamma_-=
\Omega_i(x^+)\Omega_j(x^+)=0
\ee
because $\Gamma_-$ is nilpotent. On the other hand, $\Omega_+$ does 
not satisfy these relations because it contains the combination $\Gamma_-\Gamma_+$ 
which is not nilpotent.

Since the spin connection vanishes along the transverse directions, 
it is trivial to solve the corresponding equations:
\be
\partial_i\epsilon+\Omega_i(x^+)\epsilon=0~~~~~~\longrightarrow
~~~~~~
\epsilon=\left(1-x^i\Omega_i(x^+)\right)\chi
\ee
where $\chi$ is an unconstrained spinor depending only on $x^+$.

The remaining nontrivial equation corresponds to the $+$ direction. In the 
following we will suppress the dependence on $x^+$, with the understanding that 
both $\Omega$ and $A$ are $x^+$-dependent.
\be
\partial_+\left[\left(1-x^i\Omega_i \right)\chi \right]
+{1\ov 2}A_{ij} x^j\Gamma_-\Gamma_i\chi 
+\Omega_+ \left(1-x^i\Omega_i \right)\chi =0
\ee
It is clear that the terms with different $x^i$ dependence should cancel 
separately. Thus, the equation above splits in two parts,
one of which can be used to remove from the other one terms with derivatives
acting on the spinor. The final result is
\bea
0&=&\partial_+\chi + \Omega_+ \chi 
\nn\\
0&=&-\left(\partial_+\Omega_i \right)\chi +
[\Omega_i ,\,\Omega_+ ]\chi 
+{1\ov 2}A_{ij} \Gamma_-\Gamma_j\chi ~~.
\label{split}
\eea
Being a first order differential equation, the first equation always 
has the solution
\be
\chi=e^{-\int dx^+\,\Omega_+}\rho
\ee
where $\rho$ is an unconstrained, constant spinor.

The second equation is more restrictive. 
Consider a wave solution supported by both NSNS flux  $H_{(3)}$
as well as RR fluxes which we will generically denote as 
$F\equiv\sum_pF'_{(p+1)}$. Then,  $\Omega_M$ is given by:
\be
\Omega_M=-{1\ov 8}\Gamma^{NP}H_{MNP}\otimes\sigma^3
+{e^\phi\ov 8}
\sF{}
\Gamma_M
\label{fullOmega}
\ee
where $\sF{}\equiv \sum_p \sF{}'_{(p+1)}\otimes \sigma^{l(p)}$ 
and $l(p)$ is determined from the supersymmetry transformation rules
(\ref{IIAsusy}-\ref{IIBsusy}).

Defining $\sh{}_{(2)}$ and $\sf{}_{(p)}$ as
\be
\sH{}_{(3)}\otimes\sigma^3\equiv \Gamma_-\sh{}_{(2)}
~~~~~~~~
\sF{}_{(p+1)}\otimes\sigma^{l(p)}\equiv \Gamma_-\sf{}_{(p)}
\ee
and $\sf=\sum_p\sf{}_{(p)}$, the torsion 
$\Omega_M$ decomposes into transverse and 
light-like components as
\bea
\Omega_i&=&{1\ov 8}\Gamma_-[\sh{}_{(2)},\,\Gamma_i]-
{1\ov 8}\Gamma_-\sf{}
\Gamma_i\nn\\
\Omega_+&=&-{1\ov 4}\sh_{(2)}-
{(-)^p\ov 8}\Gamma_-\Gamma_+\sf{}~~,
\label{4.15}
\eea
while $\Omega_-=0$. We also lowered the upper $+$ index on the Dirac matrices and this leads 
to the various sign differences between equations (\ref{4.15}) and 
(\ref{fullOmega}). Then, the commutator appearing in equation (\ref{split})
becomes
\bea
\!\!\!\!
[\Omega_i,\,\Omega_+]=
{1\ov 32}\Gamma_-\left[
\left(\sh{}_{(2)}^2+\sf{}^2\right)\Gamma_i
-\{\sf{},\,\sh{}_{(2)}\}\Gamma_i
+2 \sf{}\Gamma_i\sh{}_{(2)}
-2\sh{}_{(2)}\Gamma_i\sh{}_{(2)}+\Gamma_i\sh{}_{(2)}^2
\right]
\label{gencomm}
\eea

Consider now the case when the NSNS field $H_{(3)}$  and only one of the 
RR fields, $F_{(p+1)}$ are turned on, and both
have exactly one non-vanishing, constant component. Then, the first two terms 
above represent the right-hand-side of the equation of motion, while the last 
two terms give a traceless contribution to $A_{ij}$. Therefore, 
the remaining two terms must give a traceless (or vanishing)
contribution to $A_{ij}$ if the equation of motion is to be satisfied. 

Since $F_{(p+1)}$ and $H_{(3)}$ combine to form a projector in the dilatino 
variation, it is not hard to see that  $\sf$ and $\sh_{(2)}$ commute, which 
implies that the two terms we are interested in can be written as
\be
2\sf{}[\Gamma_i,\,\sh{}_{(2)}]~~.
\ee
Moreover, the vanishing dilatino variation implies that 
$\sf{}_{(p)}\chi$ and $\sh{}_{(2)}\chi$ are proportional. Therefore, the 
object above can always be written as $C_{ij}\G_-\G_j~~$, where $C_{ij}$ 
is a constant matrix. Its trace is the obstruction to constructing
a solution of the field equation with 24 supercharges and the NSNS and 
RR fluxes described above, and it vanishes.

An important question is whether any wave solution preserving more than
16 supercharges can have $x^+$-dependent form fields. If such a field existed, 
it would follow that $\partial_+\Omega_i$ in equation
(\ref{split}) is nonvanishing. Its Dirac matrix structure allows a 
contribution of $F_{(2)}$ be canceled by introducing off-diagonal entries 
of the coefficient matrix $A_{ij}$. However, the differences between the
$gl(2)$ structures of the two terms prevents this cancellation.
Thus, we conclude that all form fields must be constant.


\subsection{Solutions with  $28$ supercharges}

We now analyze the field configurations put forward in 
section \ref{secdilatino}. Of the potential solutions with 28 supercharges, 
some do not solve the Killing spinor equations. 
Those which solve it exist in type IIB and can be extended to 
include the 5-form field strength as well.

Let us begin with the type $IIB$ theory and discuss the fields in equation
(\ref{fc1}) and its S-dual version. These fields do not satisfy the 
assumptions 
introduced at the end of the previous subsection, so we must start with 
equation (\ref{gencomm}). Consider first the field configuration in 
equation (\ref{fc1}). Since $\sf$ vanishes, the 
second and third terms in (\ref{gencomm}) are absent. Furthermore,
from the previous section we know that the dilatino variation is proportional 
to $\sh{}_{(2)}$ which is
\be
\sh{}_{(2)}=f(x^+)(\Gamma_{12}+\beta\Gamma_{34}+\gamma\Gamma_{56}
+\alpha\beta\gamma\Gamma_{78})\otimes\sigma^3~~.
\label{solutie}
\ee
Thus, taking $\chi$ to be the spinors that annihilate the dilatino variation,
the only terms that survive in the second equation (\ref{split}) are
\be
0={1\ov 32}\G_-\sh_{(2)}^2\G_i\chi+{1\ov 2}A_{ij}\G_-\G_j\chi~~.
\label{ex1}
\ee
To find $A_{ij}$ it is helpful to notice that, for any choice of the index $i$ in 
equation (\ref{ex1}), passing $\G_i$ through $\sh_{(2)}$ changes the sign 
of exactly one of the four terms in $\sh_{(2)}$. Then, the fact that 
$\sh_{(2)}$ annihilates $\chi$ implies that the three terms with the sign 
unchanged can be replaced by the fourth one, whose square is proportional 
to the identity matrix. For example, for $i=1,2$ we 
have
\bea
(\sh{}_{(2)})^2\Gamma_i\chi&=&f^2\Gamma_i(-\Gamma_{12}
+\beta\Gamma_{34}+\gamma\Gamma_{56}
+\alpha\beta\gamma\Gamma_{78})^2\chi\nn\\
&=&f^2\Gamma_i(-2\Gamma_{12})^2\chi=-4f^2\Gamma_i
\label{examplecomm}
\eea
Thus, the equation (\ref{ex1}) implies that 
\be
A_{ij}={1\ov 4}\delta_{ij}f^2~~.
\label{sol1}
\ee
It is trivial to check that the equation of motion (\ref{eom++}) is satisfied.

The same analysis applies with only cosmetic changes to any of the S-duals of 
equation (\ref{fc1}). Since the Dirac matrix structure of $\sf$ and $\sh_{(2)}$ is 
identical, and $\sh_{(2)} \chi =0$, then only the first two terms in  (\ref{gencomm})
survive; for both of them the discussion above equation (\ref{examplecomm}) applies without change.

This family of S-dual solutions can be further extended to a two 
2-parameter one by including the 5-form field strength. This is possible 
because the 5-form field strength does not appear in the dilatino 
supersymmetry transformation rule. Consider the following addition to 
equation (\ref{fc1}):
\be
F_{(5)}=gdx^+\wedge (dx^1\wedge dx^2\wedge dx^3\wedge dx^4+\alpha
dx^5\wedge dx^6\wedge dx^7\wedge dx^8)
~~,~~~~
\Gamma_{-1}\chi=\alpha\chi
\ee
Under these circumstances, $\sf_{(p)}$ in equation (\ref{gencomm}) must be
replaced with $\sf{}_{(2)}+{1\ov 2}\sf{}_{(4)}$ and this leads to:
\be
0={1\ov 32}\Gamma_-(\sf{}_{(2)}+{1\ov 2}\sf{}_{(4)})^2
\Gamma_i\chi+{1\ov 2}A_{ij}\G_-\G_j\chi
\label{mod1}
\ee
where $\sf{}_{(4)}$ is given by
\be
\sf{}_{(4)}=g(\G_{1234}+\alpha\G_{5678})~~.
\ee

Since $\sf{}_{(2)}$ and $\sf{}_{(4)}$ anticommute, the equation (\ref{mod1})
becomes 
\be
0={1\ov 32}\Gamma_-\left(\sf{}_{(2)}^2+{1\ov 4}\sf{}_{(4)}^2\right)
\Gamma_i\chi+{1\ov 2}A_{ij}\G_-\G_j\chi
\label{mod1mod}
\ee
Thus, each of the two RR field strengths gives an independent contribution
to the coefficients $A_{ij}$. This shows that under certain circumstances 
plane wave solutions can be superposed without breaking supersymmetry.

The $\sf{}_{(2)}$ dependence is treated as above while the 
$\sf{}_{(4)}$ is analyzed as in the case of the maximally supersymmetric 
plane wave solution, which we now repeat for the reader's convenience.
The important observation is that for each choice of the index $i$, pushing
$\G_i$  past $\sf{}_{(4)}$ changes the relative sign between the two terms in 
$\sf{}_{(4)}$. Then, using the chirality operator, the term with changed sign
can be mapped into the one that did not. Since each of the two terms square to 
$-g^2$ we find:
\bea
\sf{}_{(4)}&=&\sf{}^1_{(4)}+\sf{}^2_{(4)}~~~~~~~~(\sf{}^I_{(4)})^2=-g^2
\nn\\
\G_-(\sf{}^1_{(4)}+\sf{}^2_{(4)})^2\G_i\chi&=&
\G_-\G_i(\sf{}^1_{(4)}-\sf{}^2_{(4)})^2\chi=4\G_-\G_i \sf{}^1_{(4)}{}^2\chi
=-4g^2\G_-\G_i \chi
\eea
Thus, the coefficients $A_{ij}$ now become the sum of the $F_3$ and $F_5$ contribution, 
and the solution is
\bea
F_{(3)}&=&f dx^+\wedge (dx^1\wedge dx^2+ \beta dx^3\wedge dx^4+\gamma
dx^5\wedge dx^6 + \alpha \beta\gamma dx^7\wedge dx^8) \nn\\
F_{(5)}&=&gdx^+\wedge (dx^1\wedge dx^2\wedge dx^3\wedge dx^4+\alpha
dx^5\wedge dx^6\wedge dx^7\wedge dx^8) \nn\\
{A}{}_{ij}&=&{1\ov 4}(f^2+{1\ov 4} g^2)\delta_{ij}~~.
\label{28f3f5}
\eea
Even though $f$ and $g$ appear in the metric only in the combination
$(f^2+{1\ov 4} g^2)$, the field strengths retain information on $f$ and $g$
separately. Using S-duality one can reconstruct the full 2-parameter family of solutions (by 
rotating $F_3$ into $H_3$ by any angle). The maximal rotation corresponds to  
solutions with only $H_3$ and $F_5$. 

To conclude this discussion, we formulate a superposition rule for wave 
solutions:

{\it Adding any two plane wave solutions with RR  fields $F_{(p+1)}$ and $F_{(q+1)}$ 
leads to a new solution. If the corresponding $f_{(p)}$ and $f_{(q)}$ anticommute, 
the common supernumerary Killing spinors are inherited by the resulting solution.}

This statement allows one to immediately decide whether the direct sum 
of two wave solutions remains a solution by just looking at the directions
covered by the various excited field strengths. The final amount of
supersymmetry is given by the number of Killing spinors common to both solutions, which can be found from the dilatino variation only.

We now turn to the other candidate solution preserving 28 supercharges (\ref{fc2}). 
The building blocks of equation (\ref{gencomm}) are in 
this case:
\be
\sh_{(2)}=f(\gamma\G_{13}-\beta\G_{24})\otimes\sigma^3~~~~~~~~
\sf_{(2)}=f(\G_{12}+\beta\gamma\G_{34})\otimes\sigma^1
\ee
Unfortunately, (\ref{fc2}) cannot source a solution
that preserves more than 20 supercharges. Indeed, $\sf^2$ contains a term of the form
$\beta\gamma\G_{1234}\otimes\id$, which cannot be canceled either by a choice of $A_{ij}$ or by introducing other fields. This further restricts the extra Killing spinors to be eigenvectors of $\G_{1234}$ with the same eigenvalue, and thus reduces them to 4. A 5-form field strength can also be added to this configuration without further reducing its supersymmetry. Solutions containing $F_{(5)}$ and the fields in (\ref{fc2}) were explored in \cite{myers} and obtained as Penrose limits of the Pilch-Warner flow \cite{warner}.

We can also analyze the possible IIA solution preserving 28 
supercharges  (\ref{IIA28-field}). The discussion is similar 
to the one above; unfortunately, these solutions do not preserve more than 20 supercharges.

The solution sourced by NSNS fluxes found in type IIB supergravity is a solution 
of the IIA theory as well. However, in the latter case it preserves only $(14,8)$ 
supercharges, because the two fermions have opposite chiralities. 

To summarize the results thus far, we have constructed in type IIB
supergravity a two-parameter family of wave solutions preserving 28 
supercharges. They are S-dual to each other and are constructed 
by adding 5-form flux to the field configurations suggested by the projector 
analysis of section \ref{secdilatino}.

\subsection{Solutions with $24$ supercharges}

$\bullet$ Deformations of solutions with 28 supercharges.

Any of the solutions discussed in the previous section and preserving 
28 supercharges can be deformed to solutions preserving only 24. Indeed, if one modifies one of the two commuting projectors in the dilatino equation $(1+M)(1+N)$, the other one is still a projector, and still annihilates half the spinors it acts upon.
Thus, all the families of solutions of the form (\ref{28f3f5})  with $\beta^2=1, \gamma^2 \neq 1$ or vice-versa preserve 24 supercharges. Adding one structure of five-form field strength can be done without paying any cost. Thus we find a two 2-parameter family of pp-wave solutions with 24 supercharges:
\bea
F_{+12}= F_{+56} =f.~~~~~ F_{+34}=\alpha F_{+78}&=& \beta f \nn\\ 
F_{+1234}=\alpha F_{+5678} &=&  g \nn\\
 A_{11}=A_{22}=A_{55}=A_{66} &=& {g^2 + 4 f^2 \over 16} \nn\\
 A_{33}=A_{44}=A_{77}=A_{88} &= &{g^2 + 4 f^2\beta^2 \over 16},
\label{f24}
\eea
together with its S-dual cousins.

The ratio $f/g$, $\beta$ and the S-duality parameter are unconstrained.
In the case $\beta=0$ we recover the Penrose-G\"uven limit of $AdS_3 \times S^3 \times T^4$ \cite{bmn}, which is also the T-dual of the solutions with 24 supercharges obtained in type IIA by \cite{c2}. If the five-form field is vanishing and $ \beta=0$, then the ``most distant'' S-dual cousin of (\ref{f24}) (involving only $H_{(3)}$) is a 24 supercharge solution of type IIA supergravity.

In chapter 3 we also discussed some projectors that preserve 
24 supercharges and cannot be extended to preserve 28. We now 
construct the supergravity solutions which realize them.

$\bullet$ Type $IIB$ supergravity

As we saw in the previous section, the field configuration (\ref{fc2}) cannot be completed to a full supergravity solution with 28 supercharges. Nevertheless, it is possible to use it for constructing solutions with 24 supercharges by truncating it to
\be
\alpha H_{+13}=F'_{+12}=f(x^+),~~~~\alpha^2=1
\ee
The dilatino variation is:
\bea
\delta\lambda&=&-{f(x^+)\ov 4}\Gamma_-
\left[\alpha\Gamma_{13}\sigma^3+
\Gamma_{12}\sigma^1
\right]\epsilon
=-{f(x^+)\ov 4}\Gamma_-\Gamma_{12}\sigma^1
\left(1+\alpha\Gamma_{23}(i\sigma^2)\right)\epsilon,
\eea
and the gravitino variation fixes the metric coefficients to
\be
A_{ij}={f^2\over 16}{\rm diag}(9,1,1,1,1,1,1,1)
\label{f3h3}
\ee
As expected, the equation of motion
\be
R_{++}=Tr A = f^2 = {1\ov 2}(H_{+13}H_+{}^{13}+F_{+12}F_+{}^{12})
\ee
is also satisfied. 

$\bullet$ Type $IIA$  supergravity

A similar solution to the one obtained above involves $F_{(4)}$ and $H_{(3)}$:
\be
{1\ov 2}H_{+12}=\alpha F_{+145}=f(x^+)~~~~~~~~\alpha^2=1,
\ee
and thus 
\be
\delta\lambda={f(x^+)\ov 2}\Gamma_-\left(
\Gamma_{12}\sigma^3+\alpha\Gamma_{145}\sigma^1\right)=
{f(x^+)\ov 2}\Gamma_-\Gamma_{12}\sigma^3\left(
1-\alpha\Gamma_{245}(i\sigma^2)\right);
\ee
the gravitino variation fixes the metric coefficients to
\be
A_{ij}={f^2\over 16}{\rm diag}(9,25,1,1,1,1,1,1),
\ee
and the equation of motion
\be
R_{++}=Tr A = 5f^2/2 = {1\ov 2}(H_{+12}H_+{}^{12}+F_{+145}F_+{}^{145})
\ee
is also satisfied.

Another type IIA solution preserving 24 supercharges can be obtained by 
combining 
$H_{(3)}$ and $F_{(2)}$:
\be
{3}F_{+1}=-{\alpha\ov 2}H_{+12}=f(x^+)~~~~~~~~~~\alpha^2=1.
\ee
The dilatino variation is given by (\ref{dilf2h3}), and the metric is given by the gravitino variation to be:
\be
A_{ij}=-{2\ov 9}{\rm diag}(121,169,1,1,1,1,1,1).
\ee
Despite these rather bizarre numbers, the equation of motion is also satisfied:
\be
 R_{++}= Tr A ={37 \over 18} ={1\ov 2} ({1\ov 3^2} + 2^2)= {1\ov 2}(F_{+1}F_+{}^1+ H_{+12}H_+{}^{12}) 
\ee

The last solution discussed in section 3 (\ref{f2f4}):
\be
3F_{+1}=F_{+234}=f(x^+)
\ee
also preserves 24 supercharges. The dilatino variation is (\ref{dilf2f4}), and the metric determined by the gravitino variation:
\be
A_{ij}= {f^2\ov 9\cdot 4}\pmatrix{4\id_4 & 0\cr  0    &\id_4 }
\ee
satisfies the equation of motion
\be
R_{++}={5\ov 9}f^2={1\ov 2}(F_{+1}^2+F_{+234}^2).
\ee
Upon lifting this solution to M-theory one obtains the maximally 
supersymmetric solution found in \cite{f2}.
We can also use the superposition principle formulated in the 
previous chapter to add to this solution the identical solution 
with fields along different directions. As we explained, the 
gravitino variation equation is satisfied if the fields anticommute, 
and the dilatino variation becomes the sum of two projectors. Thus 
the superposition solution with
\be
3F_{+1}=F_{+234}=f, ~~~ 3F_{+2}=F_{+156}=g
\ee
and the corresponding $A_{ij}$ preserves 4 supernumerary Killing spinors 
and thus has 20 supercharges.



\section{T duality}
It is interesting to explore the metrics one obtains by T-dualizing some of 
the solutions with augmented supersymmetry found in the previous 
sections. The Killing spinors that survive the T-duality transformation are those which commute
with the Killing vector defining the duality direction. Equation (4.3) implies that all
spinors depend on the transverse coordinates, therefore these directions cannot be used for our
purpose. We thus explore duality transformations along  $x^+$, which is the most interesting of 
the remaining directions.

However, as one can see from the solutions described in the previous 
chapter, all 
$A_{ij}$ giving augmented supersymmetry are positive, and 
therefore (\ref{4.1}) implies that 
$x^+$ is timelike. Unfortunately, timelike T-duality is hard to
interpret physically since it yields RR-field kinetic terms with the wrong sign
 \cite{9803259,c1}.  Thus, it can only be used as a solution-generating technique, and only for 
spacetimes with NSNS fields.

There are two ways we can circumvent this problem. The first one is to T-dualize only the 
solutions with NSNS flux. We have one such solution with 28 
supercharges, as well as
2 families of solutions with 24. The second \cite{c1} is to perform a 
coordinate 
transformation $x^- \rightarrow x^--{c \over 2} x^+$, where $c$ is 
a positive constant.
The metric (\ref{4.1}) becomes:
\be
ds^2=-2dx^+dx^- + c (dx^+)^2 - A_{ab}(x^+)z^az^b(dx^+)^2+(dz^a)^2,
\ee
and thus for any $c$ there exists a region of space where 
$x^+$ is spacelike and T-duality can be performed. The same shift can be performed for
the spacetimes containing only NSNS fields. It is rather straightforward to take any of the solutions we have and T-dualize it using the rules in \cite{t}.

As explained in the beginning of this section, not all of the original supersymmetries survive the T-duality procedure. Only those Killing spinors which are independent of $x^+$ remain Killing spinors of the new geometries. From (\ref{split}) we can see that these spinors satisfy $\Omega_+ \chi =0$. 

Unfortunately, for the solutions with 28 supercharges, $\Omega_+$ is not proportional to any projector from the dilatino variation. This is because $\sf$ and $\sh$ are no longer multiplied from the left by $\G_-$ (like in the dilatino variation), and therefore the chirality of the spinors cannot be used to combine the $\G$ matrices into products of projectors.
It is however not hard to see that, when T-dualizing the solutions with 28 supercharges, all the supernumerary Killing spinors disappear, and only 6 of the 16 annihilated by $\G_-$ remain.

PP-wave solutions with only two nonzero structures of $F_3$ or $H_3$ (preserving 24 supercharges) have more $x^+$ independent Killing spinors. Indeed, in both cases $\Omega_+$ contains one projector, and thus all the 8 supernumerary Killing spinors and 8 of the 16 regular ones survive the T-duality. The result of the duality transformation along $x^+$ is a non-pp wave solution of type IIA and 11d supergravity with 16 supercharges. 

Let us first consider a solution containing only NSNS fluxes (like the S-dual of (\ref{f24})):
\be
H_{+12}=H_{+34}=h, ~ \rightarrow ~ B_{+1}=h x^2,~~B_{+3}=h x^4, ~~H=c-{h \over 4}[(x^1)^2+(x^2)^2+(x^3)^2+(x^4)^2].
\ee
The T-dual of this geometry is
\bea
ds^2&=& {1 \over H}[(dx^+ + h x^2 d x^1+h x^4 dx^3)^2 - (dx^-)^2]+ (dx^i)^2 \nn  \\
e^{2\Phi}& =&  {1 \over |H|},~~~ B= {1 \over H} (dx^+ + h x^2 d x^1+h x^4 dx^3) \wedge dx^-,
\label{tf1}
\eea
which is exactly the metric of smeared F-strings perturbed with transverse fluxes. The solution diverges at finite distance from the origin.

Since (\ref{tf1}) only contains NSNS fields, it makes sense as a solution when $H$ is negative. The only difference is that $x^-$ becomes spacelike, $x^+$ becomes timelike, and the $B$ field switches sign. Since $|H|$ can be chosen to be nowhere vanishing (by choosing $c<0)$, this solution is regular everywhere. 

It is quite surprising that these metrics preserves 16 supercharges, and it is even more surprising that such  metrics are T-dual to that of a pp-wave.  Very similar solutions can be obtained by T-dualizing the solution with $H_3$ and 28 supercharges. In that case only 6 of the original 28 supercharges survive T-duality; however it is possible that the resulting solution preserves a larger amount of supersymmetry, of which only 6 supercharges commute with T-duality. We did not investigate this possibility.

For positive $H$ we can also T-dualize the solution with nontrivial $F_3$:
\be
F_{+12}=F_{+34}=f, ~ \rightarrow ~ C_{+1}=f x^2,~~C_{+3}=f x^4, ~~H=c-{f \over 4}[(x^1)^2+(x^2)^2+(x^3)^2+(x^4)^2],
\ee
and obtain a solution corresponding to smeared F-strings perturbed with transverse RR 2-form:
\bea
ds^2&=& {1 \over H}[(dx^+)^2 - (dx^-)^2]+ (dx^i)^2 \nn  \\
e^{2\Phi}& =&  {1 \over |H|},~~~ B_{+-}= {1 \over H},~~~C_1= f x^2,~~~C_3= f x^4
\label{f1smear}
\eea
Upon lifting this solution to M-theory we can obtain the supergravity solution of smeared M2 branes (with the harmonic function $H$), perturbed with off-diagonal components of the metric. Like the previous solutions obtained by spacelike T-duality, these solutions become divergent at finite radius.

One can T-dualize the other solutions we found and obtain geometries corresponding to F1 strings and M2 branes deformed with transverse forms. It is also possible to add to $H$ the regular harmonic function $N\over r^6$, in which case the supernumerary Killing spinors disappear, but a fraction of the regular ones survives the T-duality. One can thus obtain more realistic perturbed M2 brane solutions. 

\subsection{The AdS-CFT interpretation of the divergences}

All the solutions we found by spacelike T-duality, as well as the solutions found in \cite{c1}
have the generic property that the curvature diverges at a finite radius. Since all these solutions correspond to smeared F1 strings or M2 branes perturbed with transverse fluxes, it is possible to give them a very interesting interpretation from the point of view of the AdS-CFT correspondence.

To do this, we first add the usual harmonic function ${N\over r^6}$ to $H$. 
The metrics obtained above are still solutions, but they only have 8 
supercharges. Nevertheless, it now becomes possible to interpret them 
as near-horizon geometries of F1 strings or M2 branes perturbed with 
constant transverse $F_2$, off-diagonal metric components, or transverse $F_4$. It is quite 
straightforward\footnote{See \cite{f1d2} for the AdS-CFT analysis of the 
perturbation of F1 strings with transverse $F_2$, and \cite{m2} for the 
AdS-CFT analysis of the perturbation of M2 branes with transverse $F_4$.} 
to see that these perturbations correspond to turning on an irrelevant 
operator in the boundary theory. In the case of the M2 branes, the transverse 
perturbation with constant $F_4$ corresponds to a boundary operator of 
dimension 5, of the form $F^2\Psi\Psi$ \footnote{This can be seen from equations 
(16) and  (17) in \cite{m2}, and is similar to equation (50) in \cite{ps}}.

Since the operator is irrelevant, if one turns on a finite perturbation in 
the UV, it flows to zero in the IR. Conversely, if one turns on a finite 
perturbation in the IR, it diverges in the UV. Thus, the only solutions 
which are regular at
infinity are those with $f=0$, which is exactly what the 
solutions (\ref{f1smear}) and the ones discussed in section 7 of \cite{c1} imply.

This singularity can also be seen as coming from ``negative mass'' 
smeared M2 branes effectively created by the combination of the 
transverse 4-form (or $F_2$ and $F_6$ in the F1 string case) via the 
Chern-Simons term
of the 11d supergravity Lagrangian. When one puts enough real M2 
branes in the geometry (by adding to the harmonic function a constant 
or $N/r^6$), the supergravity is regular up to the radius where $H$ become 
zero, which is the radius where the ``negative mass'' M2 branes overtake 
the real ones. Such setups are very reminiscent of the ones where an 
enhan\c con mechanism is responsible for the removal of 
singularities \cite{enhancon}, and it would be interesting to explore if this 
is also the fate of the singularities present here.

Besides these divergent solutions we can also obtain metrics which are 
everywhere regular by adding to $H$ the function $-N/r^6$ 
and performing timelike T-duality. Of course, the wrong sign of $-N/r^6$ is 
unphysical in the original pp-wave metric, but since we are only using timelike 
T-duality for solution generating we do not worry about this. We obtain the 
metric (\ref{tf1}) with $-H = |H|={N \over r^6}+{f \over 4}\sum_i{(x^i)^2}$, which 
can again be interpreted as the near-horizon of F1 strings or M2 branes 
perturbed with off-diagonal metric components and $B_{(2)}$. 

Unlike its cousin obtained by spacelike T-duality, this solution does not diverge at finite radius. The two solutions correspond to turning on {\em different} perturbations in the IR (in one case the $B_{(2)}$  perturbation contains a timelike direction and in the other it does not). Therefore it is not surprizing that these perturbations give rise to different UV physics. 

In the regular case, the metric in the UV becomes (in the string frame)
\be
ds^2 = {f\over 2 u}[-(dt+f x^2 dx^1+f x^4 dx^3)^2 + (dx^-)^2 + du^2+ 4 u^2 d\Omega_3^2]+ \sum_{i=5}^{8}{(dx^i)^2}
\label{weird}
\ee
where $u={f\over 2} \sum_{i=1}^{4}{(x^i)^2}$. The nontrivial part of this metric is conformal to a fibration of a $\Z_2$ orbifold of a 4-plane, and does not appear to be singular. If only one structure of $H_{(3)}$ is turned on, the metric resembles that of a wave. This flow can easily be lifted to M theory, or dualized to other flows. 
Thus we obtained a nonsingular supergravity flow, starting from $AdS_4 \times S^7$ (or the near horizon F1 string metric) in the IR and ending with the geometry (\ref{weird}) in the UV. It would be very interesting to find if this
geometry has a field theory dual, and learn more about these irrelevant perturbations. Moreover, by using T and S -duality it is possible to construct similar nonsingular flows from $AdS_5 \times S^5$ in the IR to a metric similar to (\ref{weird}) in the UV. 

These types of flows are reminiscent of the one obtained by turning on the dimension 6 operator 
$B_{12}$ in the $AdS_5 \times S^5$ dual of the $\N=4$ Yang Mills theory. In that case
one also flows to a nontrivial UV geometry, which is dual to a noncommutative field theory.

\subsection{PP-waves as solution factories}

As we have seen in the beginning of this section, by T-dualizing pp-waves 
solutions with fluxes one can obtain metrics corresponding to branes 
and strings perturbed with constant fluxes. It is quite trivial to further 
use T-duality and S-duality on these solutions to
generate the solutions corresponding to other branes perturbed with 
transverse fluxes. 

However, since we consider wave backgrounds in which the form 
field strengths only depend on $x^+$ (otherwise constructing 
solutions of the equation of motion and Bianchi identities 
becomes more challenging), 
the resulting fluxes will not depend on the transverse directions, and will generically 
correspond to turning on irrelevant operators in the boundary theory. 
Thus, most of these solutions have either singularities at a finite 
distance from the origin, or very nontrivial UV completions.

Since the fluxes do not depend on transverse directions, it does not appear possible to 
obtain from the simple pp-wave ansatz the full solutions corresponding 
to perturbations of the AdS-CFT correspondence with {\it relevant} 
operators (in fact, it seems quite remarkable that pp-wave backgrounds 
are T-dual to a perturbation of the AdS-CFT duality in the first place). A 
possible direction toward obtaining these full solutions would be to go 
backward along the chain of dualities, to first obtain a more 
generic wave and then use its simple features to try to
find the full solution. 

The metrics obtained by T-duality can be easily made time-dependent. As we explained in the previous chapter, adding $x^+$ dependence to the forms and the metric removes the supernumerary Killing spinors. Nevertheless, a certain fraction of the 16 spinors annihilated by $\G_-$ (1/2 or 3/4, depending on the fluxes)  survive the T-duality. Thus, we  obtain time-dependent metrics with nontrivial fluxes, and some 
supersymmetry (8 or 12 supercharges).


\section{A general analysis \label{general}}

In this section we will prove that a wave solution with nontrivial dilatino 
variation cannot preserve more than $28$ supercharges and that the field 
configurations analyzed in sections \ref{secdilatino} and \ref{eom} are the 
only ones with this property. As discussed in the beginning of this paper, a 
systematic way of constructing all solutions with more than 16 supercharges 
is to start from 
the dilatino variation and ask for field configurations that organize it as
a product of commuting projectors. Thus:
\be
\delta\lambda=\Gamma_- M\left(\prod_{i=1}^nP_i\right)\,\epsilon
~~~~~~~~~~
P_i={1\ov 2}(1+A_i)
\label{generalsusydil}
\ee
where $M$ is some combination of Dirac matrices and $P_i$ are
a set of commuting projectors. 

An upper bound on the number of preserved supercharges translates
into an upper bound on the number of projectors that can be 
generated in the dilatino variation by the fields present in the theory.

The basic observation that will help us reach our goal is that {\it any} two 
terms in the dilatino variation must form a projector, up to a
common factor. It is easy to see that this is the case
by expanding the brackets in equation (\ref{generalsusydil}). Furthermore, 
any of these terms  has to be generated by one of the form fields 
appearing in (\ref{IIAsusy}-\ref{IIBsusy}). This implies that,
for the cases we are interested in, 
the dilaton cannot contribute to the dilatino variation. 
Indeed, after factoring out the Dirac matrix $\G_-$ which is common to all fields, 
the contribution of any component of any form field 
squares to $-1$ while the dilaton contribution squares to one. 
Similar arguments lead to the conclusion that the axion cannot
contribute either. Since the 0-form field strength cannot 
contribute to a wave solution\footnote{The equations of motion are 
not satisfied in the presence of a cosmological constant unless
form fields are allowed to have non-null nonvanishing components.}
we are left to consider $H_{(3)}$, $F_{(2)}$ and $F_{(4)}$ in 
the type IIA theory and the two three-forms of the type IIB theory.
We will now argue that it is not possible for the dilatino variation 
to contain more that two projectors besides $\G_-$

\subsection{The IIB theory}

We begin by discussing the type IIB theory. Because all spinors 
appearing in this theory have the same chirality, we can use the chirality 
operator $\Gamma_{-1}$ to rewrite a product of $m$ Dirac matrices as a 
product of $(8-m)$.

A simple inspection of the available form fields reveals that in IIB 
the prefactor $M$ in (\ref{generalsusydil}) must be a product of two 
Dirac matrices 
tensored with either $\sigma^1$ or $\sigma^3$. Then, the $gl(2)$ 
component of any of the projectors in  (\ref{generalsusydil}) is
either the identity matrix or $i\sigma^2$ depending, 
respectively, on whether only one or both types of fields are excited. 
Therefore, the Dirac matrix part of all $A_i$-s in  (\ref{generalsusydil})
must commute. 
Furthermore, they can be either products of two or four Dirac matrices.
These observations set an upper bound on the number $n$ of projectors.
In particular, there are at most three {\it independent} commuting products of two  Dirac 
matrices\footnote{Products of two Dirac matrices 
generate $SO(8)$ whose rank is four. Due to the fact that the chirality 
of both spinors is the same and that the dilatino variation is proportional 
to $\G_-$, it follows that in the dilatino variation in the type IIB theory
one of the four Cartan generators of SO(8) can be expressed in terms of 
the other three and the chirality operator $\G_{-1}$.} and only 
two {\it independent} 
commuting products of four  Dirac matrices. We will now discuss 
separately the possible constructions of projectors.

$1)$ The easiest to analyze is the case in which all $A_i$-s are built out of
products of four Dirac matrices. Since there are only two such independent
combinations, it follows that $n\le 2$ which in turn implies that there are 
at most $28$ preserved supercharges. This product of projectors, which can 
be generated using either  one of the two 3-form field strengths present in 
the theory, was  analyzed in sections \ref{secdilatino} and \ref{eom}.

$2)$ Consider next the situation when all projectors are constructed out 
of products of two  Dirac matrices. The product of projectors 
can be expanded as
\be
M\,\sum_{k=0}^n\sum_{\sigma_k\in C_n^k} \prod_{j\in\sigma_i}A_j
\ee
where $C_n^k$ is the collection of sets $\sigma_k$ of $k$ elements picked 
out of $n$. Since all $A_i$-s are different, they will commute with each other 
if and only if  no two have common Dirac matrices. The only way for this to 
come from a sum of bilinears of Dirac matrices of the type appearing in the
dilatino transformation rule, is that exactly one of the matrices building
any  $A_i$ appears in $M$. Indeed, if this were not the case, the product 
between $M$ and the corresponding $A_i$  would contain four Dirac 
matrices and this cannot be generated by one of the available 
fields\footnote{The 5-form field strength does not appear in the dilatino 
variation.}. 
Furthermore, such a term cannot be canceled using $\Gamma_{-1}$ since 
all terms in the sum above are proportional to $M$ and the use of 
$\Gamma_{-1}$ would produce terms without this property. Since, as argued above 
$M$ must be a bilinear in Dirac matrices, we can have
at most two projectors of this type in (\ref{generalsusydil}) and therefore
there are at most $28$ supercharges. Such products of projectors 
can be generated using combinations of the two 3-forms and were analyzed 
in sections \ref{secdilatino} and \ref{eom}.

$3)$ The last possibility is to have  some projectors constructed out of products 
of four  Dirac matrices while the others of products of two. The 
requirement that they commute implies that there must be an even number of common
Dirac matrices between any two $A_i$ and $A_j$. If one product of two Dirac matrices, call it $B_2$, is 
not contained in one product of four of them, call it $B_4$,
then, expanding the brackets in (\ref{generalsusydil}) implies that
we need a form field to supply a term of the type $M B_2B_4$.
But such a field does not appear in the dilatino transformation rule
unless one of the Dirac matrices appearing in $M$ also appears either
in $B_2$,  in $B_4$ or in both. Indeed, if this were the case, then $M B_2B_4$
will become a product of six $\G$-matrices and using $\G_{-1}$
can be rewritten as a product of two of them. Furthermore, similar 
arguments applied on the terms $MB_4$ implies that $M$ and
$B_4$ cannot have a common Dirac matrix. We are therefore left 
with the following possibility:
\be
\Gamma_{ab}(1+\Gamma_{bc})(1+\G_{defg})
\label{2-4}
\ee
This combination has the potential  of preserving $28$ supercharges and 
was analyzed in sections \ref{secdilatino} and \ref{eom}. 
Since there does not seem to be any obstruction to adding more projectors 
we will  attempt to do so. It is easy to see that a projector 
constructed out of four Dirac matrices that satisfies both $1)$ and $3)$ 
will anticommute with $\G_{defg}$ and thus is not allowed. We are 
thus left with the possibility
of adding a projector constructed out of two Dirac matrices. This 
will have to comply with both the restrictions of point $2)$ as well as 
with those of point $3)$. Thus, it seems possible to insert 
\be
1+\G_{ah}
\ee
where the index $h$ represents the matrix which does not already appear in 
(\ref{2-4}). Nevertheless, the three projectors are not independent because
$\G_-\G_{ah}\G_{-1}=\G_-\Gamma_{bc}\G_{defg}$ and thus the third projector 
does not lead to more preserved supersymmetry.

This concludes the analysis of the type IIB theory with the result that
any solution of the equations of motion which leads to a nontrivial dilatino
variation will preserve at most $28$ supercharges. Because the 
5-form field strength does not appear in the dilatino variation, it can be 
used to enlarge the set of fields producing the projectors discussed above.
This possibility was discussed in detail in section \ref{eom}.

\subsection{The IIA theory}

We now turn to the analysis of the type IIA theory. The discussion is 
complicated by having fields contributing different numbers of Dirac matrices
but is also simplified by the fact that we are no longer allowed  to use 
the chirality operator $\G_{-1}$ to map products of $\G$-matrices into 
each other. Indeed, any projector constructed by using $\G_{-1}$ would 
lead to enhanced supersymmetry in one sector while breaking it
in the other sector. Furthermore, in the IIA theory all fields appear in
the dilatino variation. Thus, the set of fields leading to projectors in this
variation cannot be enlarged.

As discussed before, the dilaton cannot be excited in a wave solution
with augmented supersymmetry.  
As a first step in answering the question of how many independent 
commuting projectors can appear in the dilatino variation, we first study 
if it is possible to have  two projectors. Thus, 
\be
\delta\lambda\sim M(1+A_1)(1+A_2)\epsilon  = [M+M(A_1+A_2)+MA_1A_2] \epsilon ~~.
\label{2exp}
\ee
As noticed before, each term above must be produced by 
one of the fields present in the background. We have at our disposal
products of one, two and three Dirac matrices. 

Let us now discuss case by case the the possible matrices $M$ and for each 
of them the allowed projectors $A$.

$1)$ $M$ is generated by the 2-form field strength, i.e. 
$M=\G_a\otimes  (i\sigma^2)$. The fact that $MA_i$ must be
generated by one of the fields implies that the Dirac matrix component 
of $A_i$ is constructed out of one, two or three\footnote{This last possibility 
occurs when $A$ and $M$ have one common Dirac matrix.} matrices and the 
requirement of $(1+A)$ being a projector fixes the $gl(2)$ component.
Combining everything we are left with the following possibilities: 
$\G_b\otimes g$ with $g=\id,\,\sigma^1,\,\sigma^3$, 
$\G_{bc}\otimes (i\sigma^2)$ and $\G_{abc}\otimes (i\sigma^2)$.
It is easy to see that some of these possibilities cannot be generated by 
the available fields. Indeed,  the products 
$\G_a\otimes  (i\sigma^2)\G_{bc}\otimes (i\sigma^2)$
and $\G_a\otimes  (i\sigma^2)\G_{abc}\otimes (i\sigma^2)$
have the identity matrix as $gl(2)$ component and there is no field with 
this property except for the dilaton which does not contribute any $\G$
matrix. The remaining possibility 
is $A=\G_b\otimes g$ with $g=\id,\,\sigma^1,\,\sigma^3$. 
Inserting these remaining combinations in (\ref{2exp}) we find that the 
$gl(2)$ component is fixed by the terms $MA$ to be $g=\sigma^1$.
This last possibility is nevertheless eliminated by considering 
$MA_1A_2$. Thus, we conclude that if $M$ is generated by the 2-form
field strength, the dilatino variation contains at most one projector besides $\G_-$, 
and thus no more than $24$ supercharges can exist.

$2)$ The next possibility is for $M$ to be generated
by the NSNS 3-form field strength, i.e. $M=\G_{ab}\otimes  \sigma^3$.
Then, requiring that $1+A$ is a projector, we have the following 
possibilities\footnote{We will put from 
the outset some common Dirac matrices between $A$ and $M$. This is 
due to the fact that the product $MA$ must have at most three Dirac matrices
for such a term to be generated.}:

$\bullet A=\G_c\otimes \{\id,\,\sigma^{1,3}\}$. The case with 
$\id$ and $\sigma^3$ cannot be generated due to the $gl(2)$ component while 
the other one can be generated using the 2-form field strength if 
$c=a$ or $c=b$.

$\bullet A=\G_{ac}\otimes (i\sigma^{2})$. One of the Dirac matrices that 
appear in $A$ must also appear in $M$ since otherwise there would be 
four $\G$ matrices in $MA$. It then follows that this projector cannot 
be generated due to the $gl(2)$ component of the product $MA$.

$\bullet A=\G_{acd}\otimes (i\sigma^2)$. One of the Dirac matrices that 
appear in $A$ must also appear in $M$ since otherwise there would be 
five $\G$ matrices in $MA$. Then the $gl(2)$ component of the product 
$MA$ requires that $c\ne b$ and $d\ne b$ for this projector to be generated.

$\bullet A=\G_{abcd}\otimes \{\id,\sigma^{1,3}\}$. Two of the Dirac 
matrices that 
appear in $A$ must also appear in $M$ since otherwise there would be 
four $\G$ matrices in $MA$. This in turn fixes the $gl(2)$ component of
$A$ to be the identity matrix. Such a term can be generated by the NSNS 3-form 
field strength.

$\bullet A=\G_{abcde}\otimes \{\id,\sigma^{1,3}\}$. Two of the Dirac 
matrices that 
appear in $A$ must also appear in $M$ since otherwise there would be 
more than three $\G$ matrices in $MA$. Then, the $gl(2)$ component 
prevents this term from being generated.

Thus, $A_1$ and $A_2$ must be of the type 
$\G_{acd}\otimes (i\sigma^2)$ with $c\ne b$ and $d\ne b$ (i.e. they must
have one Dirac matrix common with $M$) or of the type 
$\G_{abef}\otimes \id$. The $gl(2)$ component of $MA_1A_2$ forbids 
both $A$-s be of the second type. Thus, we have two choices. 

If both are of the 
first type then, due to the $gl(2)$ component of $MA_1A_2$ being $\sigma^3$, 
this term must be generated by the NSNS 3-form and thus must contain 
exactly two Dirac matrices. The only possibility for this to happen is if 
the Dirac matrices that are 
common between $A_i$ and $M$ are different and there is one more 
Dirac matrix common  between the two $A_i$-s. Thus, the only solution is: 
\be
\G_{ab}\otimes  \sigma^3\left(1+\G_{acd}\otimes (i\sigma^2)\right)
\left(1+\G_{bce}\otimes (i\sigma^2)\right)~~.
\label{IIAproj1}
\ee
This projector is generated by the following choice of fields:
\be
{1\ov 2}H_{+ab}=-{1\ov 2}H_{+de}=F_{+bcd}=F_{+ace}
\ee
which is none other than the field configuration discussed in equation 
(\ref{IIA28-field}).  

If the two $A_i$-s are of different type, then the $gl(2)$ component of 
$MA_1A_2$ implies that this term is generated by the 4-form field strength.
Thus, there must be another Dirac matrix common between $A_1$ and 
$A_2$ which uniquely identifies the projectors as:
\be
\G_{ab}\otimes  \sigma^3\left(1+\G_{acd}\otimes (i\sigma^2)\right)
\left(1+\G_{abce}\otimes \id\right)~~,
\label{IIAproj3}
\ee
which is just a rewriting of equation (\ref{IIAproj1}).

It is now easy to analyze the problem of adding more projectors to either 
one of equations (\ref{IIAproj1}) or (\ref{IIAproj3}). Let us discuss  
equation (\ref{IIAproj1}). If $A_3$ is of the same type as  $A_1$ and 
$A_2$, then the Dirac matrix that is common between $A_3$ and $M$ 
must be different from the ones common between $A_1$ and $M$ and 
$A_2$ and $M$. However, $M$ is constructed out of only two $\G$-matrices. 
Therefore, a third projector of the first type is forbidden. If $A_3$ is 
of the 
second type, it follows from equation (\ref{IIAproj3}) that it must have
one common Dirac matrix with $A_1$ and $A_2$ for $MA_{1,2}A_3$
to be generated. Furthermore,  this matrix cannot be common between
$A_1$ and $A_2$ because otherwise $MA_1A_2A_3$, which is 
proportional to $\sigma^3$, could not be generated. Thus, we are left with 
\be
A_3=\G_{abde}=A_1A_2
\ee
But this does not lead to an independent projector:
\be
(1+A_1)(1+A_2)(1+A_1A_2)=(1+A_1+A_2+A_1A_2)(1+A_1A_2)=
2(1+A_1)(1+A_2)
\ee
since $A_i^2=1$. 

Thus, if $M$ is generated by the NSNS 3-form field 
strength, the dilatino variation contains at most two projectors besides 
$\G_-$ and there are at most 28 preserved supercharges.

$3)$ The third and last possibility is for $M$ to be generated by the
4-form field strength, i.e. $M=\G_{abc}\otimes  \sigma^1$. 
As in the previous case, there are several possibilities 
for $A_i$\footnote{As in the previous discussions, we will put from 
the outset some common Dirac matrices between $A$ and $M$. This is 
due to the fact that the product $MA$ must have at most three Dirac matrices
for such a term to be generated.}:

$\bullet A=\G_d\otimes \{\id,\,\sigma^{1,3}\}$. There is no field that can give
this contribution. Indeed, $MA$ is a product of either four or two Dirac matrices tensored with 
$\id,\,\sigma^1$ or $(i\sigma^2)$. None of these terms can be generated
by the available fields.

$\bullet A=\G_{ad}\otimes (i\sigma^{2})$. This leads to an $MA$ with  
$\sigma^3$ as $gl(2)$ component, which requires two Dirac matrices. There 
is no choice of $d$ that can do this, and thus such an $A$ is not allowed.

$\bullet A=\G_{abd}\otimes (i\sigma^2)$. This leads again 
to $\sigma^3$ being the $gl(2)$ component of $MA$. If $d\ne c$ it also contains  
a product of two Dirac matrices. This can be generated using the NSNS 3-form field strength. 

$\bullet A=\G_{abde}\otimes \{\id,\,\sigma^{1,3}\}$. The choices $\id$ and 
$\sigma^1$ as $gl(2)$ components cannot be generated while  the choice 
$\sigma^3$ can be generated using $F_{(2)}$ if $d=c$.

$\bullet A=\G_{abdef}\otimes \{\id,\,\sigma^{1,3}\}$. The first possibility 
is viable if $d=c$, since it can be generated by the NSNS 3-form. The other 
two cases cannot be generated due to a mismatch between the $gl(2)$ 
component and the Dirac matrix component of $MA$.

$\bullet A=\G_{abdefg}\otimes (i\sigma^{2})$. The Dirac matrix component 
requires $d=c$, leading to a product of three matrices which does not 
match with the $gl(2)$ component which is $\sigma^3$. Therefore, this
combination is not allowed.

Thus, if $M=\G_{abc}\otimes  \sigma^1$ both $A_1$ and $A_2$
can be either of the type $\G_{abd}\otimes (i\sigma^2)$ with $d\ne c$, 
$\G_{abce}\otimes \sigma^{3}$ or $\G_{abcfg}\otimes \id$. By analyzing 
the six inequivalent combinations it follows that $A_1$ and $A_2$ cannot 
be of different types because of a mismatch between the $gl(2)$ component
and the number of Dirac matrices that can be generated in $MA_1A_2$.
Therefore, they must be of  the same type, which requires that $MA_1A_2$ be 
built out of three Dirac matrices since its $gl(2)$ component is $\sigma^1$.
Since $(\G_{abc})(\G_{abce})(\G_{abcf})$ contains five Dirac matrices, 
while $(\G_{abc})(\G_{abceg})(\G_{abcfh})$ contains either five or seven 
Dirac matrices, it follows that the only possibility is:
\be
\G_{abc}\otimes  \sigma^1
\left(1+\G_{abd}\otimes (i\sigma^2)\right)
\left(1+\G_{bce}\otimes (i\sigma^2)\right)~~~~~c\ne d\ne e
\label{IIAproj2}
\ee
i.e. $A_1$ and $A_2$ each have two common $\G$ matrices with $M$
and one common between themselves.
This projector is just the rewriting of (\ref{IIAproj1})

In this case there exists  $A_3=\G_{acf}$ which has the same properties as 
$A_1$ and $A_2$. However, adding it to equation (\ref{IIAproj2}) 
does not lead to enhanced supersymmetry. Indeed, the term
$MA_1A_2A_3$ will contain six Dirac matrices. Such a term can be 
generated only using $\G_{-1}$ to map it to a product of two 
$\G$-matrices, but this operation enhances supersymmetry in the 
left-handed sector while breaking it in the right-handed sector. The required
field configuration has the potential of preserving $(15,8)$ supercharges.

This concludes the analysis of the dilatino variation in type IIA theory
with the result that there exist field configurations leading to variations 
proportional to equations (\ref{IIAproj1}) and (\ref{IIAproj2}) which 
potentially preserve $28$ supercharges. Unlike the case of the IIB theory
where the 5-form does not appear in the dilatino variation, in the IIA theory 
we cannot enlarge the set of fields that lead to enhanced supersymmetry.

To summarize, we have shown that a wave solution of the supergravity 
equations of motion with a nontrivial dilatino variation preserves
at most $28$ supercharges. The candidates are given by the projectors 
analyzed in sections \ref{secdilatino} and \ref{eom}.

\section{Summary and Conclusions}

Using the relative simplicity of pp-wave geometries we have explored 
wave-like solutions of type IIA and Type IIB supergravity with
augmented amounts of supersymmetry. Making use of the chirality of 
the fermions of IIB supergravity, and expressing the dilatino variation 
as a product of commuting projectors, we found one two-parameter 
family of IIB solutions with 28 supercharges, as 
well as one IIB three-parameter family of solutions with 24 supercharges. 
We also found several individual solutions of  type IIA supergravity 
preserving  24 supercharges, as well as solutions preserving (14,8)  
supersymmetry. 

In the process of doing this, we  formulated a superposition rule for wave 
solutions, giving an easy way of testing when a direct sum of wave solutions 
still possesses enhanced supersymmetry. We also conducted a rigorous 
exploration of the possibility of constructing solutions with augmented 
supersymmetry, and concluded that  28 supercharges is the most one can 
find when the dilatino variation is nontrivial.

By T-dualizing some of our pp-wave solutions we obtained solutions 
with 16 supercharges similar to smeared F1 strings perturbed with 
transverse fluxes. After adding an extra term to the harmonic function 
we interpreted these solutions as perturbations of the AdS-CFT duality 
with irrelevant operators. This allowed us to give 
a field theoretical interpretation to the singularities some of these solutions 
generically have at finite distance from the origin. 

We also obtained an exact nonsingular IR $\rightarrow$ UV flow from 
$AdS_4 \times S^7$  to an intriguing UV geometry. Similar procedures can be
used to construct flows from $AdS_5 \times S^5$ and other near-horizon geometries.

Since our solutions are exact, the careful 
investigation of this geometries can yield further insights into the 
role of these irrelevant operators. The similarity of these flows to 
the flows to geometries dual to non-commutative field theories deserves 
further investigation, and could yield interesting physics.
 
The work presented here can be extended in several different directions.
One would be to try to realize these 
pp-wave solutions as Penrose-G\"uven limits of other supergravity 
backgrounds. This does not seem straightforward, especially for 
the solutions with 28 supercharges; their multifarious mix of fields 
makes them hard to obtain as such limits. Nevertheless, some of their 
cousin solutions (like the one with 24 supercharges and only two nonzero 
structures of $F_3$) can easily be obtained as limits of the 
$AdS_3 \times S^3$ geometry \cite{bmn}, so it is not implausible that a 
careful analysis could in the end find a ``mother background.'' If this 
background were found and it had a dual field theory, our backgrounds 
would be dual to limits of the field theory with 28 supercharges, which 
can yield interesting insights into that theory. 

Another possibility would be to look for waves whose duals 
correspond to {\em relevant} perturbations of the AdS-CFT 
correspondence. As explained in section 5, such waves would 
have $r$ dependent fields, and they would not be as simple as 
the ones discussed here. However, it is conceivable that the 
equations of motions would take a more transparent form, which 
could be more amenable to finding exact solutions.

The duality between pp-waves and 
irrelevant perturbations of the AdS-CFT correspondence could be further refined. 
As seen in 
\cite{bmn}, string theory on a pp-wave background is dual to a large 
$R$-charge sector of a field theory. After T-dualization, the 
resulting background can only be interpreted as a perturbed near-horizon 
geometry (which is dual to another filed theory) only if one adds by hand 
a term of the form ${Q\ov r^6}$ in the harmonic  function. However, this 
spoils the original duality with the large angular momentum sector of the first 
field theory. It thus appears that when $Q$ goes to zero, the dual field 
theory changes drastically, although this is not such a drastic change 
from the point of view of the supergravity. This phenomenon deserved 
further study, and might even be a link toward establishing a more direct 
relation between the field theories at the ends of the ``broken'' duality chain.

Yet another direction involves  investigating and maybe expanding the 
exact nonsingular IR to UV flows we constructed, finding possible field 
theory duals of the UV geometry, and understanding their similarity to 
the flows to non-commutative theories.

Last but not least, arguments similar to those leading to the equivalence 
of the $\N=3$  and $\N=4$
vector multiplets in 4 dimensions imply the equivalence of $\N=7$ (28 supercharges) 
and  $\N=8$ gravity multiplets. It would be interesting to see if this structure is 
preserved by interactions in the background of the waves with 28 supercharges.


{\bf Acknowledgments} We would like to thank Joe Polchinski for 
interesting and useful discussions and reading the manuscript. We also profited 
from conversations with Mariana Gra\~na and Gary Horowitz. The work 
of I.B. was supported in part by a UCSB Dissertation Fellowship, and in 
part by the NSF grant PHY00-9809(6T). The work of R.R. was supported in 
part by the DOE grant 91ER40618(3N) and in part by the NSF grant 
PHY00-9809(6T).


\end{document}